\documentclass[sigconf, nonacm]{acmart}
\usepackage[notransparent]{svg}
\usepackage{amsmath}
\begin{document}

\newbool {extended_text}
\setbool {extended_text} {true}
\newcommand\ifextended[1] {\ifbool {extended_text} {{\color {black}#1}} {}}
\newcommand\ifconference[1] {\ifbool {extended_text} {} {{\color {black}#1}}}
\newcommand\ifthen[2] {\ifbool {#1} {#2} {}}

\title {Offset-value coding in database query processing}

\author {Goetz Graefe} \orcid {0000-0003-0194-6466}
\affiliation { \institution {Google Inc}
\ifextended { \city {Madison} \state {Wisconsin} \country {USA} }
} % affiliation
\email {GoetzG@Google.com}

\author {Thanh Do} \orcid {0000-0001-9893-5725} \authornote {Work done at Google Inc.}
\affiliation { \institution {Celonis Inc}
\ifextended { \city {Madison} \state {Wisconsin} \country {USA} }
} % affiliation
\email {T.Do@Celonis.com}

\begin {abstract}
Recent work~\cite {DG-22-OVC} shows how offset-value coding speeds up database query execution, not only sorting but also duplicate removal and grouping~(aggregation) in sorted streams, order-preserving exchange~(shuffle), merge join, and more. It already saves thousands of CPUs in Google's Napa and F1~Query systems, e.g., in grouping algorithms and in log-structured merge-forests.
\newline
In order to realize the full benefit of interesting orderings, however, query execution algorithms must not only consume and exploit offset-value codes but also produce offset-value codes for the next operator in the pipeline. Our research has sought ways to produce offset-value codes without comparing successive output rows one-by-one, column-by-column. This short paper introduces a new theorem and, based on its proof and a simple corollary, describes in detail how order-preserving algorithms (from filter to merge join and even shuffle) can compute offset-value codes for their outputs. These computations are surprisingly simple and very efficient.
\end {abstract}

\maketitle

\ifextended {This paper is the extended version of an EDBT 2023 publication~\cite {GD-23-ovc-qe}.}
 
\section {Introduction}

\ifthen {true} {
% -----------------------------------------------------------------------------------------
% ZZZ: thanh attempts to address the reviewer comments
% -----------------------------------------------------------------------------------------
Tree-of-losers priority queues~\cite {Knuth-98-Sorting, Goetz-63-Sort} and offset-value coding~\cite {Conner-77-OVC, Iyer-o5-UPT-CFC} enable the most efficient sort algorithms\ifextended { for large records}. When sorting a database table with $N$ rows, the provable lower bound for row comparisons is $ log_2 (N!) \approx N log_2 (N/e) $ with $ e \approx 2.718 $. External merge sort using tree-of-losers priority queues for both run generation and merging exceeds this lower bound by only 1-2\%. Offset-value coding truncates shared row prefixes and turns prefix sizes into order-preserving surrogate keys. These surrogate keys limit the count of column value comparisons for $N$ rows with $K$ key columns to $ N \times K $~\cite {DG-22-OVC}. Importantly, there is no $ log (N) $ factor. Offset-value codes alone decide the remainder of the $ log_2 (N!) $ row comparisons. 

Recent work shows how offset-value coding complements “interesting orderings”~\cite {Selinger-79-Access} to speed up database query processing, not only sorting but also duplicate removal and grouping~(aggregation) in sorted streams, order-preserving exchange~(shuffle), merge join, and more~\cite {DG-22-OVC, DGN-22-SortAgg}. For the full benefit of interesting orderings, all execution algorithms relying on sort order must also exploit offset-value coding in their comparisons of rows and columns. Moreover, order-preserving query execution algorithms must not only consume but also produce offset-value codes, to be consumed and exploited by the next operator in the pipeline.

Nevertheless, computing offset-value codes in order-preserving query execution algorithms has not received any attention, and the only method known to-date -- comparing an operator's output row-by-row, column-by-column -- is so expensive that it renders producing offset-value across database operators much less attractive or even worthless. In contrast, new, simple techniques efficiently compute offset-value codes for order-preserving algorithms in database query execution. The new techniques do not require comparisons row-by-row, column-by-column; instead, offset-value codes in the output depend only on offset-value codes in the inputs. There is no need for additional column value comparisons beyond those required in the operation itself, e.g., column value comparisons in the merge logic of merge join. The new techniques are implemented in Google's Napa~\cite {Google-21-Napa} and F1~Query~\cite {Google-13-F1, Google-18-F1} systems.

\ifextended {The rest of the paper is structured as follows. We summarize related prior work in Section~\ref{sec:related} and explain offset-value coding and tree-of-losers priority queues in Section~\ref{sec:background}. In Section~\ref{sec:operators}, we first establish a solid foundation for efficient computation of offset-value codes by introducing a new theorem, its proof, and a pertinent corollary. Based on this theory, we then present how offset-value codes can be efficiently derived for filter, project, segmented sorting, duplicate removal, grouping, merge join (inner, outer, and semi joins), nested-loops join, order-preserving exchange, and more. We present performance results in Section~\ref{sec:eval} and finally conclude the paper in Section~\ref{sec:sum}.}
}

% -----------------------------------------------------------------------------------------
% Goetz' earlier introduction
% -----------------------------------------------------------------------------------------
\ifthen {false} {
Tree-of-losers priority queues~\cite {Knuth-98-Sorting, Goetz-63-Sort} and offset-value coding~\cite {Conner-77-OVC, Iyer-o5-UPT-CFC} enable the most efficient sort algorithms for large records.
When sorting a database table with $N$ rows, the provable lower bound for row comparisons is $ log_2 (N!) \approx N log_2 (N/e) $ with $ e \approx 2.718 $. External merge sort using tree-of-losers priority queues for both run generation and merging exceeds this lower bound by only 1-2\%. Offset-value coding truncates shared row prefixes and turns prefix sizes into order-preserving surrogate keys. These surrogate keys limit the count of column value comparisons for $N$ rows with $K$ key columns to $ N \times K $~\cite {DG-22-OVC}. Importantly, there is no $ log (N) $ factor here. Offset-value codes alone decide the remainder of the $ log_2 (N!) $ row comparisons. Equally importantly, offset-value coding can avoid column value comparisons throughout sort-based query processing, not only in merge sort but also, for example, in merge join and in-stream duplicate removal~\cite {DG-22-OVC}.

\ifextended {
As offset-value coding provides a surrogate key for fast row comparisons in sorted streams and in merge sort, it provides benefits like hash values in hash aggregation, hash join, etc. Both in memory and when spilling, hash-based algorithms are more similar to distribution sort~\cite {Isaac-56-DistrSort} than merge sort. Neither their algorithmic logic nor their inputs, intermediate results, and outputs can benefit from offset-value coding and the implicitly enabled compression by prefix truncation.

Another difference between offset-value codes and hash values is that, due to the danger of hash collisions, hash values can only prove that two rows or their keys differ, whereas offset-value codes can also prove that two rows or their keys are equal. An offset equal to the key size always indicates a duplicate key.

Finally, the computation of a hash value, e.g., in duplicate removal, requires accessing and decoding all columns in a row or its key, whereas the computation of an offset-value code touches only those columns truly required to compare and sort a pair of rows (and tree-of-losers priority queues ensure the minimal count of row comparisons, i.e., $log_2 (N!)$ for $N$ rows). In fact, row comparisons produce offset-value codes as a side-effect, practically for free.
} % if extended text

Sort-based query processing is particularly efficient if multiple operations focus on the same column or set of columns, well known as “interesting orderings”~\cite {Selinger-79-Access} in relational query optimization. For the full benefit of interesting orderings, all execution algorithms relying on sort order must also exploit offset-value coding to speed up their comparisons of rows and columns. Moreover, order-preserving query execution algorithms must not only consume but also produce offset-value codes, to be consumed and exploited by the next operator in the pipeline. An obvious but rather inefficient method, yet the only method to-date, compares successive output rows column-by-column. Put differently, the remaining gap in prior work is an efficient technique that computes offset-value codes for the output of merge join and all other order-preserving algorithms in database query execution. 
Our research, after introducing a new theorem, its proof, and a pertinent corollary, fills this gap for filter, project, segmented sorting, duplicate removal, grouping, merge join (inner, outer, and semi joins), nested-loops join (lookup join), order-preserving exchange (merging shuffle), and more.

\ifextended {Nonetheless, their core logic is so concise that mainframes implement leaf-to-root passes in tree-of-losers priority queues with the UPT “update tree” instruction and offset-value coding with the CFC “compare and form codeword” instruction~\cite {IBM-88-ISA-370}. Iyer~\cite {Iyer-o5-UPT-CFC} sums up that “Together, the UPT and CFC instructions do the bulk of sorting in IBM’s commercial DBMS DB2 running on z/Architecture processors.” \ifextended {This implies that tree-of-losers priority queues and offset-value coding do the bulk of sorting.} The present paper aims to let offset-value codes do the bulk of predicate evaluation in sort-based query algorithms for joining, grouping, and more.}

} % -----------------------------------------------------------------------------------------

\section {Related prior work} \label {sec:related}

The context of our work are Google's Napa~\cite {Google-21-Napa} and F1~Query~\cite {Google-13-F1, Google-18-F1} systems. Napa is a data warehouse that maintains thousands of materialized views in log-structured merge-forests~\cite {OCG-96-LSM}. F1~Query is a federated query processing platform that executes SQL queries over tables in various Google storage systems such as Spanner~\cite {Google-13-Spanner}, BigTable~\cite {Google-08-BigTable}, and Napa. Both Napa and F1~Query employ sort order for efficient data access and data manipulation.

Pioneering work on sorting established the benefits of tree-of-losers priority queues and of offset-value coding, both for run generation and external merge steps~\cite {Goetz-63-Sort, Conner-77-OVC, Knuth-98-Sorting, Iyer-o5-UPT-CFC}. Pioneering work in the database field established the value of sort-based query execution, notably merge join but also duplicate removal and grouping~\cite {Blasgen-Eswaran-77, Selinger-79-Access, Epstein-79-Aggregates}. Surveys on database query evaluation~\cite {G-93-Survey, G-06-Sorting} cover sorting and offset-value coding but fail to recognize offset-value coding as a significant opportunity throughout sort-based query execution. Recent work~\cite {DG-22-OVC, DGN-22-SortAgg} fills that gap but fails to offer efficient derivation of offset-value codes for output rows. The present short paper fills this remaining gap.

\begin{table*}[t]
\small 
\caption {Offset-value codes in a sorted file or stream.} \label {table:ovc}
\centering
\begin{tabular}{|r|r|r|r| r|r|r| r|r|r|}
\hline
% Table header
\multicolumn{4}{|c|}{Rows and their} &
\multicolumn{3}{c|}{Descending OVC} &
\multicolumn{3}{c|}{Ascending OVC} \\
\cline{5-10}
\multicolumn{4}{|c|}{column values} &
$\textit{offset}$ & $domain - value $ & OVC & $\textit{arity} - \textit{offset}$ & $value$ & OVC \\
\hline
\textbf{5} & 7 & 3 & 9 & 0 & 95 & 95 & 4 & 5 & 405 \\
\hline
5 & 7 & 3 & \textbf{12} & 3 & 88 & 388 & 1 & 12 & 112 \\
\hline
5 & \textbf{8} & 4 & 6 & 1 & 92 & 192 & 3 & 8 & 308 \\
\hline
5 & \textbf{9} & 2 & 7 & 1 & 91 & 191 & 3 & 9 & 309 \\
\hline
5 & 9 & 2 & 7 & 4 & - & 400 & 0 & - & 0 \\
\hline
5 & 9 & \textbf{3} & 4 & 2 & 97 & 297 & 2 & 3 & 203 \\
\hline
5 & 9 & 3 & \textbf{7} & 3 & 93 & 393 & 1 & 7 & 107 \\
\hline
\end{tabular}
\end{table*}

\section {Background: offset-value coding and tree-of-losers priority queues} \label {sec:background}

A tree-of-losers priority queue~\cite {Goetz-63-Sort, Knuth-98-Sorting}, also known as a tournament tree, embeds a balanced binary tree in an array, with the tree’s unary root in array slot 0. It is efficient due to leaf-to-root passes with one comparison per tree level; root-to-leaf passes with two comparisons per tree level are not required.
As in a sports tournament where each round of competition eliminates one contestant,
the principal rules are that (i) two candidate key values compete at each binary node in the tree and (ii) after a comparison of two candidates, the loser remains in the node and the winner becomes a candidate in the next tree level.
Thus, in a priority queue with $N$ entries, a new overall winner reaches the root after $ log_2 (N) $ comparisons.

When merging runs, a fixed pair of runs competes at each leaf node. Run generation merges “sorted” runs of a single row each. Run generation by replacement selection can try to extract longer sorted runs from the unsorted input: one additional comparison per input row doubles the expected run size, halves the run count, and saves one comparison per row during merging.

\begin {figure}
\centering \includegraphics [width=0.72\columnwidth] {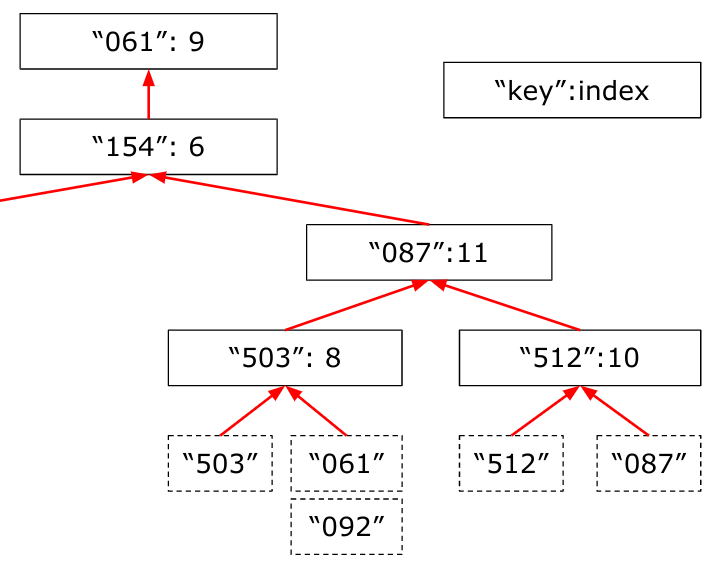}
\caption {Strings in a tree-of-losers priority queue.} \label {fig:pq-char}
\end {figure}

Figure~\ref {fig:pq-char}, adapted from~\cite {Knuth-98-Sorting}, shows a tree-of-losers priority queue merging $F = 12$ sorted runs. The left half with runs 0-7 is cropped from Figure~\ref {fig:pq-char}. The dashed boxes represent the current candidates from runs to be merged; the solid boxes are tree nodes in a tree-of-losers priority queue. The example node in the top-right corner explains the numbers in each node.

The overall smallest key value, “061”, is in the tree’s root node. Interestingly, key value “154” is above key value “087”. After “154” emerged as the winner from the left subtree and “061” from the right subtree, “154” was the loser in the “final match” of this “tournament” and was left behind. The runner-up of the right subtree, “087”, had to remain within that subtree.

The next step in the merge logic moves key value “061” from the tree root to the merge output. The following step gets a new key value from input 9, the origin of key value “061”. This successor key starts a new leaf-to-root pass at the the same leaf. This leaf-to-root pass bubbles the next lowest key value to the root in $log_2 (F)$ steps: If that next key value is “092”, it wins over “503” but loses to “087” and is left behind; then “087” wins over “154” and reaches the root.

\ifextended {
With only leaf-to-root passes, run generation and merging with tree-of-losers priority queues guarantee near-optimal comparison counts. With excellent expected and worst-case run-time complexity yet very limited implementation complexity, some hardware supports tree-of-losers priority queues, e.g., the UPT “update tree” instruction of IBM’s 370- and z-series mainframes~\cite {IBM-88-ISA-370, Iyer-o5-UPT-CFC}.
} % if extended text

Offset-value coding~\cite {Conner-77-OVC} captures one row’s key value relative to another key that is earlier in the sort sequence. Offset-value codes are set after comparisons. A loser's new offset is the position where the keys first differ, e.g., a column index; the value is the loser's data value at that offset. Equivalently, the offset is the size of the shared prefix. For example, a duplicate key shares the entire key and hence has an offset equal to the key size.

Table~\ref {table:ovc} shows the derivation of descending and ascending offset-value codes in a stream of rows in ascending order on all columns. Each key is encoded relative to its predecessor. With four sort columns, the arity of the sort key is 4; the domain of each column is 1…99. Descending offset-value codes take the actual offset but the negative of the column value. Ascending offset-value codes take the negative offset but the actual column value. Table~\ref {table:ovc} ignores that small key domains permit encoding multiple key columns together.
\ifextended {
IBM’s CFC “compare and form codeword” instruction supports offset-value coding for descending normalized keys using blocks of bytes as values and counts of blocks as offsets~\cite {IBM-88-ISA-370, Iyer-o5-UPT-CFC}.
} % if extended text

If two rows and their key values $A$ and $B$ are encoded relative to the same key $C$, and if the offsets of $A$ and $B$ differ, then the one with the higher offset is earlier in the sort sequence. Otherwise, if the two data values at that offset differ, then these data values decide the comparison. Otherwise, additional data values in $A$ and $B$ must be compared. With offsets and values combined as shown in Table~\ref {table:ovc}, often a single integer comparison can sort two keys $A$ and $B$ encoded relative to the same key $C$.

If $A$ sorts lower (earlier) than $B$, then $B$ is the loser in this comparison and can be encoded relative to the winner $A$. If offset-value codes decided the comparison, the offset-value code of $B$ relative to $C$ is also the offset-value code of $B$ relative to $A$. Otherwise, the offset for $B$ increases by the count of column comparisons required to determine winner and loser, and the value is taken at that new offset. Similar rules apply if $A$ is the loser in the comparison.

\ifextended {
\begin {table}
\small 
\caption {Offset-value code decisions and adjustment.} \label {table:ovc-adj}
\centering
\begin {tabular} {|c|c|c|p{0.75cm}|p{0.75cm}|c|}
\hline
% Table header
 & \multicolumn{2}{|c|}{} & \multicolumn{3}{c|}{Offset-value codes} \\
\cline {4-6}
{Case} & \multicolumn{2}{|c|}{Keys} & \multicolumn{2}{c|}{to base (3,4,2,5)} & {loser to winner} \\
\hline
1 & {3,5,8,2} & {3,4,6,1} & 305 & 206 & 305 \\
\hline
2 & {3,4,3,8} & {3,4,9,1} & 203 & 209 & 209 \\
\hline
3 & {3,7,4,7} & {3,7,4,9} & 307 & 307 & 109 \\
\hline
\end {tabular}
\end {table}
}

\ifextended {
Table~\ref {table:ovc-adj} shows shows pairs of key values encoded relative to a shared base key. Offsets decide the comparison in case 1, values at the shared offset decide case 2, and additional column value comparisons decide case 3. The loser requires a new offset-value code only in case 3.
}

In a tree-of-losers priority queue with offset-value coding, each tree node was a loser to the local winner and the local offset-value code is relative to the local winner. Along the overall winner’s leaf-to-root path, all key values are coded relative to its key value.

\begin {figure}
\centering \includegraphics [width=0.72\columnwidth] {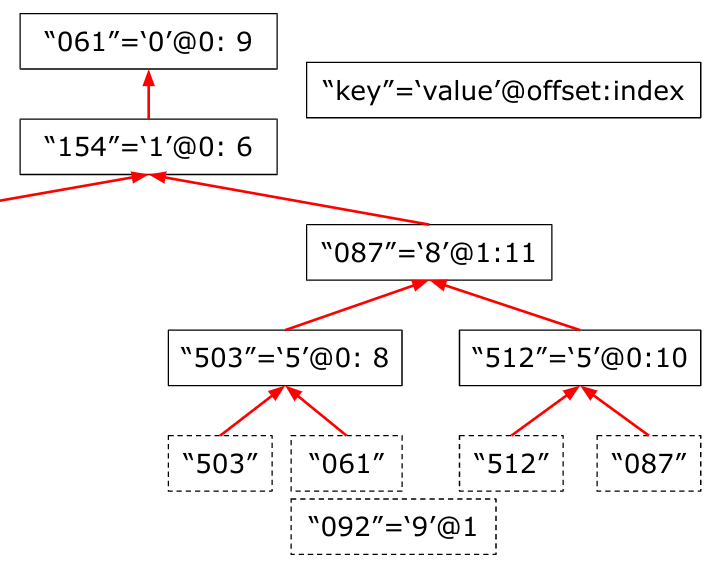}
\caption {Offsets and values in a tree-of-losers priority queue.} \label {fig:pq-ovc}
\end {figure}

Figure~\ref {fig:pq-ovc} shows the same key values as Figure~\ref {fig:pq-char} but adds offset-value codes. In the figure, each node indicates an offset and a character at that offset. Although an offset-value code for an ascending sort order requires negating either the offset or the value, this is not shown in Figure~\ref {fig:pq-ovc}. In an implementation, each node in the tree-of-losers priority queue contains only an offset-value code and an index; strings remain in the input buffers reachable because the index values are run identifiers.

Recall that input runs are encoded with prefixes truncated. For example, “092” is encoded relative to its predecessor “061”. After “061” moves from the tree root to the merge output, “092” starts a leaf-to-root pass. Keys “503”, “087”, and “154” on this path are all encoded relative to the prior overall winner “061”. Offset~1 is earlier in the sort order than offset~0; therefore, “092” is earlier than “503”. “503” stays in place and “092” moves up. There, offset~1 equals offset~1, but data value~`8' is less than data value~`9'; therefore, “092” remains behind as “087” moves up. Finally, “087” wins over “154” and reaches the root.

In this leaf-to-root pass, offset-value codes decide all three comparisons, two by offsets and one by data values encoded within offset-value codes. Not a single string comparison is required and not a single offset-value code needs re-calculation.

\ifextended {
\begin {figure}
\centering \includegraphics [width=0.85\columnwidth] {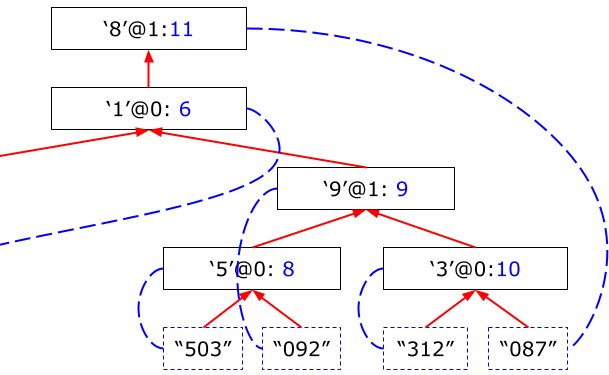}
\caption {Pointers in a tree-of-losers priority queue.} \label {fig:pq-ptrs}
\end {figure}

Figure~\ref {fig:pq-ptrs} shows the tree-of-losers priority queue of Figure~\ref {fig:pq-ovc} a bit closer to its actual implementation. The string values or database rows remain in the input buffers, while each node in a tree-of-losers priority queue holds only an offset-value code and a run identifier. The latter serves as a pointer to a specific merge input and thus to a string value or database row. Thus, nodes in the tree implementing a tree-of-losers priority queue are small and of fixed size.
} % if extended text

Merging sorted runs replaces a winner with its successor in the same merge input. With merge inputs' fixed assignment to leaf nodes, a successor retraces the leaf-to-root path of the prior overall winner. Since this successor as well as all keys on its leaf-to-root path are encoded relative to the prior overall winner, offset-value coding applies to all comparisons in a tree-of-losers priority queue.

Offset-value codes decide many comparisons in a tree-of-losers priority queue. Column value comparisons are required only if two rows have equal offset-value codes, with column comparisons starting at the offset. After such a row comparison is decided, the count of column value comparisons is added to the loser’s offset.

With $K$ sort columns, the sum of all offset increments is limited to $K$ in each row; in an input with $N$ rows, the sum of all increments and thus the count of all column value comparisons are limited to $ N \times K $. Importantly, there is no $ log (N) $ multiplier. Thus, tree-of-losers priority queue and offset-value coding guarantee that the effort for column value comparisons is linear in the count of rows and in the count of sort columns, quite like the effort for computing hash values in hash-based query execution.

Comparisons of offset-value codes are free if they are subsumed in other algorithm activities and data structures. In quicksort, for example, the inner-most loop not only compares key values but also looping indexes: when the loops from the left and the right meet, the current partitioning step is complete. In priority queues, the inner-most loop compares key values only after testing whether there even are key values. During queue construction, some entries might not have been filled yet; after the end of some merge inputs, some queue entries no longer have valid keys; and during run generation by merging many single-row runs, there practically is only queue build-up and tear-down.

A tree-of-losers priority queue for run generation by replacement selection may hold key values $ - \infty $ and $ + \infty $ as early and late fences, valid keys assigned to the current run, and valid keys for the next run. These cases need some indicator field in each queue entry, but they require only $2$ bits. Even with multiple early and late fences, e.g., one for each merge input~\cite {G-06-Sorting}, $30$ or $62$ bits remain and can hold an offset-value code. A single comparison instruction can test whether the two relevant keys are valid and compare their offset-value codes: if two rows have the same 32- or 64-bit value, then they are both valid (neither early nor late fences), they go to the same output run, they differ from their shared base row (an earlier winner) at the same offset, they have the same value at that offset, and the next step must compare further columns. As the comparison of offset-value codes is already complete when the indicator fields have been inspected for administration of the inner-most loop, offset-value code comparisons are free.

\ifextended {This design also reduces CPU cache faults. If a tree-of-losers priority queue requires 8~bytes per entry, a L1~cache can retain a priority queue (an array) of 512 or 1,024 entries. The data records may or may not fit in a lower-level cache but offset-value codes can decide many comparisons without cache fault. Mini-runs of this size remain in memory until merged (with fan-in 512 or 1,024) to form initial runs on external storage~\cite {Baer-89-Quicksort, Friend-56-Sorting, Nyberg-95-AlphaSort}.}

Offset-value coding can speed up not only merge sort but also order-preserving exchange (merging shuffle), merge join, set operations such as intersection, and more. For example, in a query like “select…, count~(distinct~…) group~by…”, the sort can detect duplicate rows by offsets equal to the column count and, after the sort, in-stream aggregation can detect group boundaries by offsets smaller than the grouping key.

\section {Offset-value coding in relational query execution operators} \label {sec:operators}

What has been missing to-date are simple and efficient techniques for deriving offset-value codes in the output of database operators other than merge sort. Surprisingly, a new theorem and its corollary enable the wanted solution with no additional column value comparisons beyond those required for computing output rows.
\ifextended {The new theorem also implies two prior theorems about offset-value codes in the context of merge sort.}

\textbf {Definitions}: For key values $A$ and $B$,
let $ pre(A,B) \geq 0 $ be the length of their maximal shared prefix;
let $ val(A,i) $ with $ i \geq 0 $ be the data at offset $i$ within key value $A$;”
let $ val(A,B) = val(B, pre(A,B)) $ be the first difference in key value $B$ relative to key value $A$;
let $ A < B $ mean that $A$ sorts lower (earlier) than $B$; and
let $ ovc(A,B) $ with $ A < B $ be the offset-value code of key value $B$ relative to key value $A$,
computed from $ pre(A,B) $ and $ val(A,B) $ as in Table~\ref {table:ovc}.

\textbf {Proposition}: For key values $ A < B < C $ with $ A \neq B $ or $ B \neq C $,
$ ovc(A,B) \neq ovc(B,C) $.

Proof (by contradiction): $ ovc(A,B) = ovc(B,C) $ would imply that
$C$ has the same data value as $B$ at the same offset, e.g., column index,
but this would violate the definition of offset-value codes, which requires the maximal shared prefix.

Examples: Table~\ref {table:ovc} shows no successive equal offset-value codes.

\textbf {Theorem}: For key values $ A < B < C $,
in ascending offset-value coding $ ovc(A,C) = max(ovc(A,B), ovc(B,C) ) $ and
in descending offset-value coding $ ovc(A,C) = min(ovc(A,B), ovc(B,C) ) $.

Proof (for ascending offset-value codes in an ascending sort, in three cases by the lengths of maximal shared prefixes):
(i)~If $ pre(A,B) > pre(B,C) $, then $ pre(A,C) = pre(B,C) $, $ val(A,C) = val(B,C) $, and $ ovc(A,C) = ovc(B,C) $. With $ ovc(A,B) < ovc(B,C) $, the theorem holds.
(ii)~Otherwise, if $ pre(A,B) < pre(B,C) $, then $ pre(A,C) = pre(A,B) $, $ val(A,C) = val(A,B) $, and $ ovc(A,C) = ovc(A,B) $. With $ ovc(A,B) > ovc(B,C) $, the theorem holds.
(iii)~Otherwise, $ pre(A,B) = pre(B,C) $ and, by the lengths of maximal shared prefixes, $ val(A,B) < val(B,C) $. Thus, $ pre(A,C) = pre(B,C) $, $ val(A,C) = val(B,C) $, and $ ovc(A,C) = ovc(B,C) $. With $ ovc(A,B) < ovc(B,C) $, the theorem holds.

Examples: Case~(i) in the proof applies to the first three rows in Table~\ref {table:ovc}. If the second row were removed, then the offset-value codes of the third row would change in neither ascending nor descending offset-value coding.
As an example of case~(ii), if the second-to-last row were removed in Table~\ref {table:ovc}, the offset-value codes of the last row would be those of the removed row.
As an example of case~(iii), if the third row were removed in Table~\ref {table:ovc}, the offset-value codes of the fourth row would remain unchanged.

\ifextended {
\textbf {Corollary} (Iyer's “unequal code theorem”~\cite {Iyer-o5-UPT-CFC}): For key values $ A < B < C $, if $ ovc(A,B) < ovc(A,C) $, then $ ovc(B,C) = ovc(A,C) $.

Proof: The corollary’s premise implies $ ovc(A,C) \neq ovc(A,B) $; the theorem above then requires that $ ovc(A,C) = ovc(B,C) $.

Implication: If offset-value codes relative to base A decide the comparison between key values B and C, then the loser's offset-value code relative to the winner is the same as its offset-value code relative to the old base A. There is no need to compute a new offset-value code for the loser relative to the winner.

\textbf {Corollary} (Iyer's “equal code theorem”~\cite {Iyer-o5-UPT-CFC}): For key values $ A < B < C $, if $ ovc(A,B) = ovc(A,C) $, then $ ovc(B,C) < ovc(A,C) $.

Proof: The corollary’s premise plus the proposition implies that $ ovc(A,C) = ovc(A,B) \neq ovc(B,C) $; the theorem above then implies that $ ovc(B,C) < ovc(A,C) $.

Implication: If offset-value codes relative to base A cannot decide the comparison between key values B and C, then their difference must be located (and column value comparisons should start) past the shared prefix and value.
} % if extended

\textbf {Corollary} (the new “filter theorem”): Our new theorem above extends to multiple intermediate keys: for a sorted list of key values $ X_{0} < X_{1} < \dots < X_{n-1} < X_{n} $ and ascending offset-value coding, $ ovc(X_{0}, X_{n}) = max_{i = 1 \dots n} ovc(X_{i-1}, X_{i}) $.

Proof (sketch): By repeated application of the theorem.

Implication: When a filter drops rows from a sorted stream, simple and efficient integer calculations can derive offset-value codes for the output from offset-value codes of the input.

\begin{table}
\small
\caption {Offset-value codes after a filter.} \label {table:filter}
\centering
\begin{tabular}{|r|r|r|r| r|r|r|}
\hline
\multicolumn{4}{|c|}{Rows and their} & \multicolumn{3}{c|}{Ascending OVC} \\
\cline{5-7}
\multicolumn{4}{|c|}{column values} & $\textit{arity} - \textit{offset}$ & $value$ & OVC \\
\hline
\textbf{5} & 7 & 3 & 9 & 4 & 5 & 405 \\
\hline
5 & \textbf{9} & 3 & 7 & 3 & 9 & 309 \\
\hline
\end{tabular}
\end{table}

\subsection {Filter} \label {sec:filter}

A filter with a predicate computes offset-value codes for its output by directly applying the “filter theorem” above: an output row’s offset-value code is (in ascending encoding) the maximum of its offset-value code in the input and of the offset-value codes of rows that failed the filter predicate since the prior output row. The same applies \textit {mutatis mutandis} for descending offset-value coding and for offset-value coding using byte offsets within normalized keys.

Table~\ref {table:filter} illustrates the calculation for ascending offset-value codes with the data of Table~\ref {table:ovc}, assuming that only the first input row and the last input row satisfy the filter predicate.

\subsection {Projection}

Projection, i.e., removal of input columns as well as calculation of new columns from existing columns (all within a single row), typically does not change the sort order. “Relationally pure” projection, however, includes removal of duplicate rows, which of course might change the sequence of rows (see Section~\ref {sec:dup}).

If all columns in the sort key survive the projection, offset-value codes in the output are the same as in the input. If not, the offset must be limited to the prefix (column count) that survives.

\subsection {Segmented sorting} \label {sec:segm_sort}

Segmented query execution means that a single stream is divided into segments, one after another, that can be processed one at a time. A typical example is a stream sorted on (A,~B) but required sorted on (A,~C) – one can either sort the entire stream on (A,~C) or one can segment the input on distinct values of (A) and sort each segment only on (C). In this example, A,~B, and~C can be individual columns or lists of columns.

To segment a sorted stream with offset-value codes, inspection of these code values suffices: an offset smaller than the segmentation key indicates a segment boundary. All other offsets must be cut to the size of the segmentation key, to be extended again by the sort within each segment. In the example above, all offsets within a segment are cut to the size of (A); the associated value is the first column value of (C). Sorting a segment on (C) refines these offset-value codes in the usual way to reflect the sort order on (A,~C). For a more concrete example using Table~\ref {table:ovc}, segmenting on the first two columns does not require any column value comparisons; an offset of less than two (and corresponding ranges of offset-value codes) indicates a boundary between segments.

\subsection {Duplicate removal} \label {sec:dup}

In a sorted stream with offset-value codes, duplicate removal suppresses input rows with offsets equal to the arity (count of columns). In Table~\ref {table:ovc}, for example, an offset of 4 (and corresponding offset-value codes) indicates a duplicate row. All other rows, i.e., the output rows, retain their offset-value codes from the input. In the duplicate-free output, no row has an offset equal to the arity. The same applies \textit {mutatis mutandis} if the sort key is reduced based on functional dependencies~\cite {Simmen-96-Order-Optn}.

\subsection {Grouping and aggregation}

In a stream with offset-value codes sorted on a “group~by” list, grouping aggregates input rows with offsets equal to or larger than the “group~by” list. In the aggregation output, no row has an offset equal to or larger than the “group~by” list. The output rows retain the offset-value codes of the first row in each group of input rows. In Table~\ref {table:ovc}, for example, grouping on the first two columns can use offset-value codes similarly to segmentation (see Section~\ref {sec:segm_sort}).

\subsection {Pivoting} \label {sec:pivot}

Pivoting turns rows into columns, e.g., from (year, month, sales) to (year, january\_sales... december\_sales). In many aspects, including the set of useful algorithms, pivoting is like grouping and aggregation. This applies in particular to the benefit of offset-value codes in the input and the calculation of offset-value codes in the output.

\subsection {Merge join} \label {sec:mj}

The logic of merge join is similar to an external merge sort; hence, it can exploit offset-value codes in its two sorted inputs. Minor variations of merge inner join can provide all join types as well as set operations such as intersection. The merge logic supports cross-table equality predicates, e.g., a primary key and a foreign key; a subsequent filter can enforce other cross-table predicates (see Section~\ref {sec:filter}, with caveats for semi joins and outer joins).

Semi joins (SQL “exists” sub-queries) and anti semi joins (SQL “not exists” sub-queries) select rows from one input based on a join predicate rather than an filter predicate within a row (see Section~\ref {sec:filter}). Nonetheless, the rule for setting offset-value codes in the output is the same as given in the “filter theorem” above, just like the derivation of Table~\ref {table:filter} from Table~\ref {table:ovc} as discussed in Section~\ref {sec:filter}.

Inner joins are similar: the offset-value codes of one input are preserved in the output, rows without match affect the offset-value code of the next row with a match, and rows with duplicate matches must have offset-value codes for duplicate keys.

\ifextended {
A left outer join preserves the offset-value codes of the left input. The same logic as for inner join applies to a left row with multiple matches. This also works \textit {mutatis mutandis} for right outer joins.

The sort order of full outer joins is unusual because join columns from both inputs can have null values in the output. A virtual column can be equal to both join columns for output rows that belong to the inner join; equal to the left join column for rows of the left anti semi join, i.e., the left outer join minus the inner join; and \textit {mutatis mutandis} for right outer joins. The approach using a virtual column also works for one-sided outer joins.
} % if extended text

\ifextended {
Among set operations, intersection proceeds mostly like an inner join, union like a full outer join, and difference like an anti semi join. Sort-based multi-set (bag) operations benefit from grouping on the input side (collapsing duplicate rows to a single row with a counter) and expansion on the output side. Query optimization can model the representation of duplicate rows (multiple copies, single copy with counter, multiple copies each with a counter) as a physical property, similar to sort order and partitioning.
The same applies to the presence of offset-value codes in a stream of intermediate results, although it seems that offset-value codes are rather beneficial in any sorted stream, except of course the input of a sort operation (but also see Section~\ref {sec:segm_sort} above).
} % if extended text

In summary, the merge join logic is similar to a merge step in an external merge sort: it might require comparisons of column values but it can compute offset-value codes for the output without additional column values comparisons.

\subsection {Nested-loops join, look-up join} \label {sec:nlj}

In the “filter theorem” above, it does not matter whether input rows fail a single-table predicate in a filter, a two-table predicate in a semi join, or a many-table predicate in nested iteration. Thus, the “filter theorem” above applies here just as much as in Sections~\ref {sec:filter} and~\ref {sec:mj}.
Nested-loops join or lookup join can be order-preserving. Note that there is no requirement that the join predicate is an equality predicate. \ifbool {extended_text} {Like most implementations of lookup join and index nested-loops join, we} {We} ignore here right semi join, right anti semi join, right outer join, and full outer join, which leaves left semi join, left anti semi join, inner join, and left outer join; we assume that the left input is the outer input and the right input is the inner input.

\ifextended {
If each result from the inner input is also sorted (on any of its columns) and includes offset-value codes, the output rows of inner join and left outer join benefit from offset-value codes of matching inner rows, with the offset incremented by the size of the outer sort key. In a many-to-many join and in the case of duplicate outer join key values and multiple matching inner rows, each inner row joins all outer rows before processing the next inner row. In other words, maximum offsets in the output’s offset-value codes require that the roles of outer and inner loops are reversed within each many-to-many match.
}

\ifextended {
\subsection {Order-preserving in-memory hash join}

Hash-join preserves the sort order of its probe input if the build input and its hash table fit in memory. This condition can be guaranteed at compile-time if the stored table is smaller than memory or if execution-time policies guarantee it, e.g., by extremely large memory allocation or by an adaptive degree of parallelism with additional memory attached to each execution thread.

In those cases, the hash table is much like an unsorted version of a database index in index nested-loops join. This includes rows with and without match in inner, semi, and outer joins.
}

\subsection {Order-preserving shuffle}

An order-preserving one-to-many “splitting” shuffle resembles a filter with respect to each output partition (see Section~\ref {sec:filter}), because each output stream is a selection from the overall input stream. An order-preserving many-to-one “merging” shuffle requires the standard merge logic, very similar to a merge step in an external merge sort. A tree-of-losers priority queue can exploit offset-value codes in the input and produce new offset-value codes in the output.
\ifextended {An order-preserving many-to-many shuffle is usually not recommended due to its danger, however small, of deadlock among producer and consumer threads~\cite {G-93-Survey}. If used, its effects on sort order and offset-value codes are similar to a sequence of many-to-one and one-to-many shuffle operations.}

\subsection {Ordered scans}
Data access is a source of offset-value codes as important as sorting. All sorted scans can produce offset-value codes.

Column storage is often sorted with the leading key columns compressed by run-length encoding. Fortunately, as described in recent work~\cite {DG-22-OVC}, such scans can produce row-by-row offset-value codes without sorting and even without any column value accesses or column value comparisons. Thus, these scans can provide offset-value codes practically for free.

Traditional b-trees readily support sorted scans. Page-wide prefix compression~\cite {Bayer-77-Prefix} gives offset-value coding a head start; compression within index leaves by next-neighbor difference, e.g., in Shore~\cite {Carey-DeWitt-94-Shore}, provides offset-value codes practically for free.

Today’s ubiquitous log-structured merge-forests and stepped-merge trees~\cite {JNS-97-Stepped, OCG-96-LSM} support a sorted scan per partition, which can include offset-value codes; a merge of such scans benefits from offset-value codes and merge logic using a tree-of-losers priority queue readily produces offset-value codes for the merge output.

In non-unique secondary indexes, lists of row identifiers are usually sorted and compressed using one of the above compression techniques and thus can deliver such lists with offset-value codes. Range queries need to merge lists of row identifiers; again, the merge logic consumes, benefits from, and produces offset-value codes. Multi-dimensional b-tree access, e.g., MDAM~\cite {LJB-95-MDAM}, similarly merges sorted lists of row identifiers. Sorted lists of row identifiers are similarly useful for index intersection and index join, i.e., “covering” a query in “index-only retrieval” with multiple secondary indexes of the same table~\cite {GBC-98-Teams}.

\subsection {Summary of new techniques}

In the sort-based and order-preserving query execution algorithms considered above, calculation of new offset-value codes in the operations’ output can be simple and efficient. Far from requiring row-by-row column-by-column data comparisons, offset-value codes in the output depend only on offset-value codes in the inputs, as proven by a new theorem and a corollary. There is no need for additional column value comparisons beyond those performed in the operation itself, e.g., column value comparisons in the merge logic of merge join. Ordered storage structures, e.g., b-trees, can preserve the effort for comparisons spent during index creation -– they can do so by storing offset-value codes explicitly, by prefix truncation (encoding each record relative to its immediate predecessor), or by run-length encoding of leading sort columns. Scans over either format can readily produce offset-value codes and, with the new techniques of this paper, sort-based operations can easily carry them forward, i.e., produce offset-value codes in their output derived from offset-value codes in their input.

\ifextended {
\section {Implementation Experience}

This section briefly summarizes how offset-value coding is used in F1~Query~\cite {Google-13-F1, Google-18-F1}. At a high level, F1~Query employs offset-value coding in its sorting logic and takes advantage of offset-value codes in intermediate results (e.g., spilled files) and across relational operators to further improve query performance.

The F1 sort operator uses external merge sort with tree-of-losers priority queues and offset-value coding for both run generation and merging. Each tree entry encodes offset and value bits in an unsigned 64-bit integer offset-value code. Invalid key values, e.g., after a merge input is exhausted, are also folded into this integer. In a sense, each comparison of offset-value codes is folded into a test whether or not a run is exhausted; thus, the comparison of offset-value codes is practically free.

Offset-value codes for rows in sorted runs are a byproduct of run generation. These offset-value codes later improve the efficiency of merging. For example, the merge logic inspects the offset-value code of the next row from the winner's input. If this next key value has an offset equal to the number of key columns, the row goes directly to the output buffer, bypassing the merge logic entirely.

Since many sort-based operators (e.g., in-stream aggregation, merge join, analytic functions, and order-preserving exchange) can leverage offset-value codes in their inputs, F1~Query supports carrying offset-value codes between operators. An artificial column for offset-value codes is introduced during query planning for order-producing physical operators. Order-producing relational operators use the logic described in this paper to produce offset-value codes in its output. The sorting techniques described in this paper are also employed in Napa~\cite {Google-21-Napa}.
} % if extended text

\section {Performance evaluation} \label {sec:eval}

This section tests two claims or hypotheses about offset-value coding, sort-based query processing, and interesting orderings:
\begin {enumerate}
\item Offset-value coding can speed up external merge sort and also its consumers, e.g., in-stream aggregation, merge join, segmentation, and order-preserving (merging) exchange. 
\item Offset-value coding together with interesting orderings can be more efficient than hash-based query plans.
\end {enumerate}

Some readers may think these claims are obviously true, but we nonetheless wish to dispel any conceivable remaining doubt. These hypotheses do not claim universally superior performance, merely new performance advantages for sort-based query processing due to offset-value coding in contexts beyond merge sort. \ifconference {Due to limited space, we refer readers to~\cite {GD-22-ovc-qe-arxiv} for more information on the implementation of offset-value coding in Napa and F1~Query.}

In order to focus on order-based operators and offset-value coding, all experiments use a single execution thread.
The hardware is an engineering workstation; more detailed specifications are omitted on purpose.
% do not include: The workstation has a 12-core Intel Xeon CPU 3.5~GHZ and 10~GB of RAM.
Each experiment starts with a warm cache, i.e., input data pre-fetched into memory.
The measurements below come from Google's public benchmark library~\cite {gbench}.
Test data are synthetic yet similar to the actual data in our daily production web analysis with many rows and many key columns. Each key column is an 8-byte integer with only a few distinct values.

\begin {figure}
\centering \includegraphics [width=0.75\columnwidth] {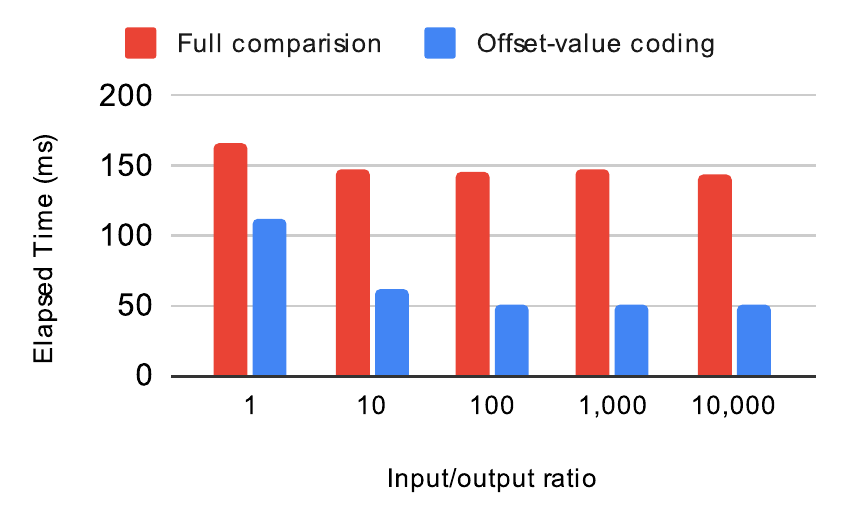}
\caption {Group boundaries from offset-value codes.} \label {fig:hyp1}
\end {figure}

Figure~\ref {fig:hyp1}, in a test of hypothesis 1, shows the effort of offset-value coding in in-stream aggregation. For fast detection of group boundaries, the operation exploits the offset-value codes from a preceding sort operation. This is compared to using full comparisons of multiple key columns. The count of input rows is 1,000,000. The ratio of input rows and output rows varies. For example, a ratio of 1 indicates all input rows are distinct (i.e., all groups have size 1) and a ratio of 100 indicates that on average 100 input rows contribute to each output row. All queries of the type “select…, count~(distinct…) group~by…” require a two-step process. Figure~\ref {fig:hyp1} shows that, within the sorted output, testing the offset against the count of grouping columns is much faster than full comparisons of multiple key columns. Thus, offset-value codes benefit not only sorting but also other query execution algorithms.

\begin {figure}
\centering \includegraphics [width=0.75\columnwidth] {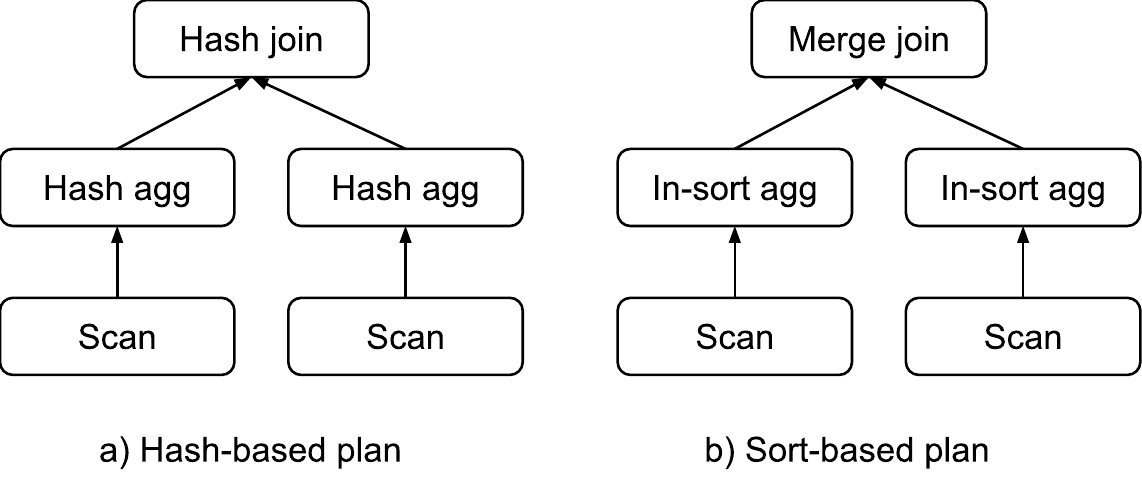}
\caption {Query plans for an “intersect distinct” query.} \label {fig:hyp2-plan}
\end {figure}

Figure~\ref {fig:hyp2-plan} shows two query evaluation plans, one hash-based and the other one sort-based, for a simple SQL query computing the intersection of two tables, e.g., “select B from T1 intersect select B from T2”. If column B is not a primary key in tables T1 and T2, correct execution requires duplicate removal plus a join algorithm. The query plans for “intersect all” would be practically the same, counting duplicates rather than removing them; and similarly for set differences using “except” and “except all” syntax. Index intersection for "and" predicates uses the same kinds of query plans, as does index join~\cite {GBC-98-Teams} when emulating columnar storage in data warehouses with traditional indexes (in the ideal case, compressed). % do not ~\cite {G-07-Columnar}

In the hash-based plan, there are three blocking operators: two hash aggregation operators for duplicate removal and a hash join for set intersection. In contrast, the sort-based plan has only two blocking operators: both are in-sort aggregation operators for duplicate removal. The merge join computing the intersection exploits not only interesting orderings but also offset-value codes in the output of in-sort aggregation. Conceivably, advanced techniques could reduce the counts of blocking operators in both sort- and hash-based query plans, e.g., integrated hashing~\cite {G-94-Volcano, GBC-98-Teams}, groupjoin~\cite {MN-11-Groupjoin}, or co-group-by instead of join~\cite {DGN-22-SortAgg}.

\begin {figure}
\centering \includegraphics [width=0.75\columnwidth] {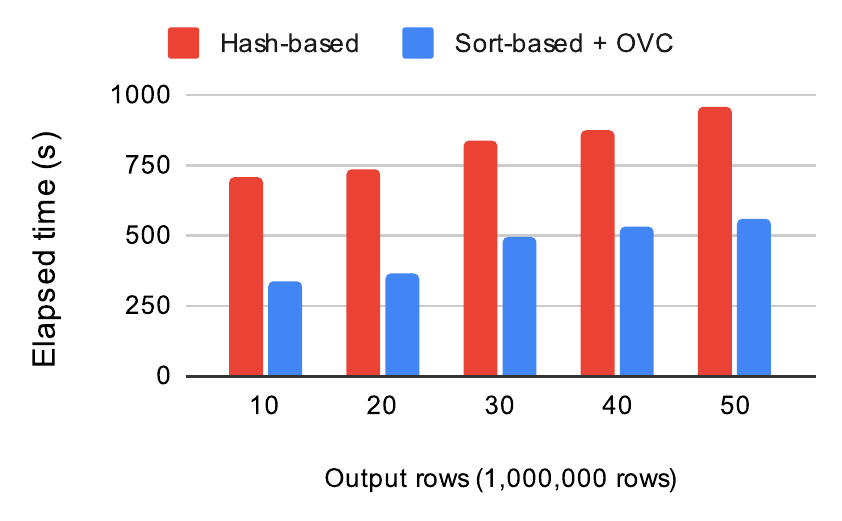}
\caption {Performance of “intersect distinct” query plans.} \label {fig:hyp2}
\end {figure}

Figure~\ref {fig:hyp2} shows the performance of the plans in Figure~\ref {fig:hyp2-plan}. Each input table has 100,000,000 rows; each blocking operator's memory holds 10,000,000 rows. In a hash-based plan, duplicate removal and join spill to temporary storage such that many rows are spilled twice.
In contrast, the sort-based plan spills each input row only once. Thus, interesting orderings cut the spilling effort in half. In addition, offset-value codes from the in-sort aggregation operators speed up row comparisons in the merge join.

In summary, while the queries in these experiments are simple and common, the measurements support our two claims about the benefits of offset-value coding in database query processing.

\section {Summary and conclusions} \label {sec:sum}

Computing offset-value codes in order-preserving query execution algorithms has not received any attention to-date for two reasons. First, new research has only recently widened the scope of offset-value coding from merge sort to all sort-based query execution algorithms and query plans, including complex join and grouping queries. Second, the only method known to-date -- comparing an operator's output row-by-row, column-by-column -- is so expensive that it thwarts all benefit of offset-value codes in query execution.

A new theorem\ifextended {~and one of its corollaries} enables an alternative method for computing or propagating offset-value codes. Beyond its use here, this new theorem already enables efficient maintenance of offset-value codes in near-optimal sorting with tree-of-losers priority queues~\cite {Graefe-23-PQ-OVC} and in b-trees with prefix truncation (during key deletion)~\cite {GD-23-storage}.

With this solid foundation, calculating offset-value codes for operator output in database query execution is surprisingly simple and very efficient. Importantly, there is no need for any column value comparisons beyond those required to compute output rows. Offset-value coding throughout sort-based query execution takes “interesting orderings” to their full potential, not only in query planning but also in plan execution. In binary operations, e.g., merge join and intersection, offset-value codes decide many or most row comparisons, whereas in unary operations, e.g., duplicate removal and grouping, offset-value codes decide all row comparisons.

\ifextended {
In hash-based query execution, a single compiled-in integer comparison of hash values can determine that two rows or their key values are different. In sort-based query execution, a single compiled-in integer comparison of offset-value codes can do the same; in addition, it can determine that two rows or their keys are equal.
Hash-based query execution requires accessing $ N \times K $ column values just for the hash function; each duplicate key adds $K$ column value comparisons. In contrast, sort-based query execution accesses only those columns required to differentiate rows or keys.
In an extreme case with a unique first column, the entire operation accesses not $ N \times K $ but only $N$ column values, each only once to prime offset-value codes in a tree-of-losers priority queue.
} % if extended text

Offset-value coding already saves thousands of CPUs in Google's Napa and F1~Query systems, because ingestion (run generation), compaction (merging), and query processing in log-structured merge-forests rely heavily on sorting and merging. We anticipate ever more savings in the future as sorting and merging are essential in many data processing pipelines and in many storage formats.

In conclusion, if storage structures are kept sorted on column values, if query optimization considers interesting orderings, and if query execution fully employs offset-value coding plus some techniques from earlier work~\cite {DG-22-OVC, G-11-g-join}, then sort-based query processing can consistently be at least as efficient as hash-based query processing. Hash-partitioning remains recommended for parallel and distributed query execution until future innovation ensures efficient, well-balanced, and failure-tolerant range-partitioning.

\begin{acks}
We thank Herald Kllapi and the reviewers for their helpful suggestions.
\end{acks}

\clearpage
\bibliographystyle{ACM-Reference-Format}
\bibliography{main}
\end{document}
\endinput